# Topological crossovers in the forced folding of self-avoiding matter

Alexander S. Balankin[1,2], Daniel Morales Matamoros[2], Ernesto Pineda León[1],

Antonio Horta Rangel[3], Miguel Ángel Martínez Cruz[1], Didier Samayoa Ochoa[1]

[1] Grupo Mecánica Fractal, Instituto Politécnico Nacional, México D.F., México, 07738

[2] Instituto Mexicano del Petróleo, Eje Lázaro Cárdenas Norte, México D.F., México

**Abstract**

We study the scaling properties of forced folding of thin materials of different geometry. The scaling relations implying the topological crossovers from the folding of three-dimensional plates to the folding of two-dimensional sheets and further to the packing of one-dimensional strings are derived for elastic and plastic manifolds. These topological crossovers in the folding of plastic manifolds were observed in experiments with predominantly plastic aluminum strips of different geometry. Elasto-plastic materials, such as paper sheets during the (fast) folding under increasing confinement force, are expected to obey the scaling force-diameter relation derived for elastic manifolds. However, in experiments with paper strips of different geometry we observed the crossover from packing of one-dimensional strings to folding two dimensional sheets only, because the fractal dimension of the set of folded elasto-plastic sheets is the thickness dependent due to the strain relaxation after a confinement force is withdrawn.

# 1. Introduction

Folding of thin materials is of noteworthy importance to many branches of science and industry [1-20], ranging from the microscopic level, such as the motion of molecules in protein folding [21-30], to the macroscopic level, such as the paper crumpling [31-50] and fault-related folding in multilayered rocks [51-54]. Folded materials are an important member of the enormous class of soft condensed matter systems [55-60]. The folding configurations may be regular, such as the artistic origami [61-63], or random, such as of the hand crumpled paper balls [31-50], highly folded and disordered S-Mo-S layers [64], two-dimensional graphite oxide membranes [65, 66], and folded carbon nanosheets [67]. Accordingly, the problems of folding and unfolding have been studied in the theoretical, computational, and experimental domains.

From the experimental point of view, randomly folded materials are ill-defined systems, because the folding procedures appear quite haphazard [68]. Nevertheless, the experiments with randomly folded matter are rather well reproducible (see [6, 31-60] and references therein), because of the topology and self-avoiding interactions are two most important physical factors when dealing with folding of thin matter [68, 69]. The relevant property of thin matter is that its stretching rigidity is much more than the bending rigidity. As a result, the metric, which determines distances between points on the surface, remains unchanged after folding deformations [49]. So, formally, a folding of self-avoiding matter is a continuum of isometric embeddings of the $d$-dimensional manifold in the n-dimensional space (see [8, 10, 70, 71] and references therein). Accordingly, the folding configurations are very strongly constrained by the topological (Euclidean) dimensions of the folded manifold ($d$) and the embedding space ($n \geq d$) (see [72-81] and references therein).

Specifically, from the energetic balance follows that the characteristic size of the equilibrium folded configuration ($R$) is related to the manifold size ($L$) as $R \propto L^{\nu_d}$, where the scaling exponent $\nu_d$ is a function of $d$ and $n$ [5, 72-76]. Therefore, it is expected that a set of $d$-manifolds of different mass ($M$) folded in n-dimensional space obeys a fractal scaling law [72, 73, 82, 83]:

$$M \propto R^{D_d(n)} f(X), \qquad (1)$$

where $D_d(n) = d/\nu_d(n)$ is the fractal dimension of the set of folded manifolds and $f(X)$ is a function of geometric constraint $X$ determined by the manifold geometry. For manifolds of thickness ($h$), width ($W \geq h$), and length ($L \geq W$) a natural form of the dimensionless geometric constraint is

$$X = LW^{\omega}h^{\eta}, \qquad (2)$$

where the scaling exponents should satisfy the obvious equality

$$1 + \eta + \omega = 0. \qquad (3)$$

For manifolds with negligible thickness ($h$) and bending rigidity, such as polymer molecules and proteins, the Flory approximation for elastic and self-avoidance energies leads to the following relationship [84]:

$$D_d(n) = \frac{d(2+n)}{2+d}, \qquad (4)$$

which agrees with results of Monte Carlo simulations [72-77, 85-88].

The universal topological properties of folded configurations of flexible polymer chains ($d=1$) in a good solvent are perfectly described by the model of self-avoiding walks (SAWs) on a regular lattice [89]. The scaling properties of SAWs on regular lattices have been studied in detail both analytically [90, 91] and in computer simulations [86-88]. For example, in the three-dimensional space ($n=3$) one finds within the frame of the field-theoretical renormalization group approach $D_1(3) = 1.69 \pm 0.01$ [90], close to the prediction with the Flory-type approximation (4).

Tethered membranes, sometimes termed polymerized or crystalline membranes, are the natural generalization of linear polymer chains to intrinsically two-dimensional structures [3, 92]. They possess in-plane elasticity as well as bending rigidity and are characterized by broken translational invariance in the plane and fixed connectivity resulting from relatively strong bonding [4]. Monte Carlo simulations of flexible self-avoiding surfaces with free boundaries give $D_2(3) = 2.5 \pm 0.03$ [93] in agreement with Eq. (4). Some physical realizations of tethered membranes, such as graphitic oxide membranes, have a crumpled conformational structure with the fractal dimension of approximately 2.5 [3, 94].

Furthermore, in the Monte Carlo simulations of flexible self-avoiding strips of width $W \gg h$ and length $L > W$, Kantor, *at al.* [72-75] have observed a crossover from self-avoiding polymers to tethered surfaces for which Flory theory predicts

$$R \propto \left(WL^3\right)^{1/(d+2)} \tag{5}$$

for $d \leq 4$ [72]. Then, for three-dimensional embedding we have the fractal scaling relation

$$M = \rho h W L \propto W L \propto R^{5/2} \left(\frac{W}{L}\right)^{1/2}, \qquad (6)$$

where $\rho$ is the material mass density. Notice that the scaling relation (6) coincides with the scaling relation (1), where the geometric constraint (2) is $X = W/L$ and the scaling function behaves as

$$f(X) \propto X^\theta, \qquad (7)$$

with the scaling exponent

$$\theta = \frac{D_2(3)}{D_1(3)} - 1. \qquad (8)$$

Accordingly, Eq. (6) interpolates between Flory predictions for one-dimensional strings ( $M \propto L \propto R^{5/3}$ , $h << W = const << L$ ) and two-dimensional membranes ( $M \propto L^2 \propto R^{2.5}$, $W = L >> h = const$ ).

It should be pointed out that the Flory approximation describes only the equilibrium configurations of randomly folded manifolds with negligible bending rigidity [3, 95]. In contrast to this, in the case of forced folding of materials with finite bending rigidity, the folded states are essentially non-equilibrium, because the most part of folding energy is accumulated in the network of crumpling creases (see [1, 2, 5-7, 96] and references therein). The analytical considerations and lattice simulations of the forced folding suggest that the fractal dimension of d-manifolds with finite bending rigidity folded in the space $n > 2d$ should coincide with the topological dimension of embedding space [78], *i.e.*,

$$D_d(n \geq 2d) = n . \tag{9}$$

in a good agreement with the experiments with metallic wires folded in two-dimensional space (see [97-99]).

Furthermore, numerical simulations of forced folding with a coarse-grained model of triangulated self-avoiding surfaces with bending and stretching elasticity [5] suggest that the characteristic size of folded configuration $R$ scales with the hydrostatic confinement force $F$ as

$$\frac{R}{L} \propto \left(\frac{YL^2}{\kappa}\right)^\beta \left(\frac{\kappa}{FL}\right)^\alpha , \tag{10}$$

where $h$ and $L$ are the thickness and edge size of square sheet ($h \ll L$), $Y$ is the two-dimensional Young modulus, $\kappa$ is the bending rigidity, $\gamma = YL^2/\kappa$ is the dimensionless Föppl–von Kármán number, and $\beta$ and $\alpha$ are the scaling exponents [5].

From (10) follows that a set of square sheets of different size $L$ and fixed thickness ($h = const \ll L$) folded by the same force ($F = const$) obeys a fractal scaling law

$$M \propto L^2 \propto R^{D_2(3)} \tag{11}$$

with the fractal dimension

$$D_2(3) = \frac{2}{1+2\beta-\alpha} \ . \tag{12}$$

Moreover, numerical simulations [5] suggest the universal values of the exponents $\beta$ and $\alpha$. Specifically, for self-avoiding elastic membranes it was found that $\alpha = 0.25$, $\beta \approx 0.06$, and $D_2(3) = 2.3$, while for phantom sheets with finite bending rigidity $\alpha = 3/8$, $\beta = 1/16$, and $D_2(3) = 8/3$ [5].

While equation (10) was proposed in [5] for the folding of elastic manifolds, experimental studies [83, 100] suggest that this scaling relation is also valid for the forced folding of predominantly plastic materials. For instance, in experiments with predominantly plastic aluminum foils it was found $\beta = 0.04 \pm 0.005$, $\alpha = 0.21 \pm 0.02$, and $D_2(3) = 2.30 \pm 0.01$ [83], while the authors of [100] report $\alpha = 0.196 \pm 008$ and $D_2(3) = 2.4$. Notice that in both experimental works [83, 100] it was found that the scaling exponents are independent of the thickness and initial size of the foil, but the experimental exponents are slightly differ from the universal exponents found in the numerical simulations of elastic manifolds. Moreover, the authors of [100] found a well different thickness and size independent exponent $\alpha = 0.065 \pm 0.002$ for high-density polyethylene films and thus they suggest that the force scaling exponent is not universal, at least for predominantly plastic materials. Early, the effect of plastic deformations on the mechanics of densely folded media was studied by Bevilacqua [79] within a geometric approach. He pointed out that the fractal dimension of randomly folded sheet should depend on the material ductility. So, taking into account the relationship (12), one can expect that the force scaling exponent $\alpha$ also depends on the material ductility in agreement with experiments [100].

In the last years, many experimental studies have been carried out with randomly folded wires [97-99, 101-104] and thin sheets of different materials with different thickness [6, 43-48, 57-59, 67, 100, 105-108]. Early, in the comment to the work [72], Gomes and Vasconselos [109] have noted that the paper strips of width $W$ and length $L > W$ folded by hands do not satisfy the scaling behavior (6). More recently, it was shown that the fractal dimension $D_2(3)$ of a set of folded elasto-plastic sheets is material dependent, because of the strain relaxation after the confinement force ($F = const$) is withdrawn [45]. Furthermore, it was found that folded elasto-plastic sheets are characterized by two different fractal dimensions – the universal local dimension of the internal configuration of folded matter ($D_l = 2.64 \pm 0.05$) and the material dependent global fractal dimension $D_2(3)$ of the set of balls folded from sheets ($h = const$) of different size $L$ [46]. The effect of sheet thickness on the scaling properties of randomly folded plastic materials was studied by Bevilacqua [80] and later in the works [83, 100]. However, as far as we know, there were no efforts to study the crossover behavior in forced folding of self-avoiding matter. Accordingly, this work is devoted to the systematic studies of scaling properties of sets of balls folded from strips of different geometry characterized by relations between strip thickness ($h$), width ($W >> h$), and length ($L \geq W$).

This paper is organized as follows. In Section 2, we derive the scaling relations describing the topological crossovers in the forced folding of ideally elastic and ideally plastic manifolds with the finite bending rigidity. Section 3 is devoted to the details of experiments performed in this work. Results of experiments with predominantly plastic aluminum foils are discussed in the Section 4. In Section 5, we present the experimental results for hand folded paper strips of different geometry, the scaling properties of which are dependent on the strain relaxation after the folding force is withdrawn. The summary of experimental results and the force-diameter scaling relation implying the topological crossovers are given in the Section 6.

## 2. Scaling analysis of the forced folding

Taking into account the dependence of bending rigidity on the sheet thickness, the scaling behavior (10) may be rewritten in the form

$$R \propto L^{2/D_2(3)} h^{\phi}, \text{ when } F = const, \qquad (13)$$

where the functional expression of the thickness scaling exponent $\phi$ in terms of the scaling exponents $\beta$ and $\alpha$ is determined by the constitutive properties of thin plates. Specifically, for an elastic plate $\kappa \propto Eh^3$ [110], where $E$ is the thickness independent three-dimensional Young modulus, and so

$$\phi = 3\alpha - 2\beta. \qquad (14)$$

Accordingly, a set of square sheet of the same thickness and size $L \gg h = const$, can be treated as a set of two-dimensional manifolds obeying the fractal behavior (11, 12), whereas a set of elastic plates with the same thickness to size ratio ($h/L = const$), can be treated as a set of three-dimensional manifolds obeying fractal behavior

$$M \propto hL^2 \propto L^3(h/L) \propto R^{D_3(3)} \;, \qquad (15)$$

where the fractal dimension $D_3(3)$ is determined by the force scaling exponent as

$$D_3(3) = \frac{3}{1+2\alpha} \;. \qquad (16)$$

Using the universal values of $\alpha = 0.25$ and $\beta \approx 0.06$ suggested in [5] for self-avoiding elastic manifolds, from Eqs. (14) and (16) one obtains $\phi \approx 2/3$ and $D_3(3) = 2$, respectively.

On the other hand, in the flexible elastic layers with the bending rigidity $\kappa \propto Yh^2$, where $Y$ is the thickness independent two-dimensional Young modulus [111], also obey scaling relations (13) and (15) but with different scaling exponents

$$\phi = 2(\alpha - \beta) \qquad (17)$$

and

$$D_3(3) = \frac{3}{1+\alpha} \;, \qquad (18)$$

respectively. Taking into account that the value of the force of scaling exponent $\alpha = 0.25$ obtained in the numerical simulations [5] does not depend on the thickness dependence of the bending rigidity, from (18) follows that for flexible elastic layers $D_3(3) = 2.4$, while the fractal dimension $D_2(3)$ defined by Eq. (12) may differ from the universal value $D_2(3) = 2.3$ obtained in [5] for elastic plates with $\kappa \propto Eh^3$, because the exponent $\beta$ for flexible layers may be different than this for elastic plates.

Notice that in both cases, the scaling relation (10) can be rewritten in the form (1) with the scaling function (7), where the geometric constraint (2) is $X = h/L$ and the scaling exponent is defined as follows

$$\theta = 2\left(\frac{D_3(3)}{D_2(3)} - 1\right). \tag{19}$$

Therefore, the scaling relation (10) implies a topological crossover from the folding of two-dimensional elastic sheets ($h = const \ll L$) to the crumpling of three-dimensional elastic plates ($h/L = const$).

Further, taking into account that a set of elastic strings ($h = const \leq W = const \ll L$) with the finite bending rigidity folded in the three-dimensional space is expected to obey a fractal behavior

$$M \propto L \propto R^{D_1(3)} \tag{20}$$

with the fractal dimension $D_1(3) = 3$ determined by Eq. (9), one can expect that for thin rectangle strips $h = const < W \leq L$ exist a topological crossover analogous to this in the case of flexible membranes. This crossover can be described by the scaling relation (1) with the scaling function (7) of the geometric constraint $X = W/L$ and the scaling exponent defined by Eq. (8).

Through the dimensional analysis, we obtain the following general scaling relation

$$M = \rho hWL \propto R^{D_3(3)} \left( \frac{Lh^\eta}{W^{1+\eta}} \right)^\vartheta, \qquad (21)$$

where

$$\eta = \frac{2D_1(3)[D_3(3) - D_2(3)]}{D_2(3)[D_1(3) - D_3(3)]}, \qquad (22)$$

and

$$\vartheta = \frac{D_1(3) - D_3(3)}{D_1(3)}, \qquad (23)$$

which describes the topological crossovers from the scaling behavior (20) of randomly packed one dimensional strings $h = const \leq W = const << L$) to the scaling behavior (11) of randomly folded two-dimensional sheets ($h = const << W = L$) and further to the scaling behavior (15) of randomly crumpled three-dimensional plates ($h/L = const$, $W = L$). For ideally elastic manifolds the fractal dimensions $D_d(3)$ are expected to be universal and so, the scaling exponents $\eta$ and $\vartheta$ are also universal. Specifically, for elastic plates with the bending rigidity $\kappa \propto Eh^3$, Eqs. (22), (23) predict $\eta = -0.78$ and $\vartheta = 1/3$, whereas for flexible elastic layers with the bending rigidity $\kappa \propto Yh^2$, Eqs. (22) and (23) give $\eta = 1/D_2(3)$ and $\vartheta = 1/5$, respectively.

In the case of ideally plastic materials, the folding work is dissipated due to plastic deformations, such that for a set of plates made from the same material the dissipated energy $w \propto Kh^2$, where $K$ is a constant depending on the physical properties of the material (see Ref. [80]). So, using the dimensional analysis technique, for plastic materials we can rewrite the scaling relation (10) in the form

$$\frac{R}{L} \propto \left(\frac{L}{h}\right)^{2\beta} \left(\frac{Kh^2}{FL}\right)^{\alpha} \qquad (24)$$

where $\beta$ and $\alpha$ are the material dependent scaling exponents (see [45, 100]). Accordingly, randomly folded plastic sheets of different thickness are expected to obey the scaling relation (13) with the material dependent scaling exponent (17). Experimentally, for aluminum foils of different thickness it was found the value $\phi = 0.35 \pm 0.01$ consistent with the value $\phi = 0.31 \pm 0.03$ given by Eq. (15) with the experimental values of $\alpha = 0.196 \pm 008$ [100] and $\beta = 0.04 \pm 0.008$ [83].

Furthermore, a set of plastic sheets of the same thickness folded by the same force ($F = const$) is expected to obey the fractal scaling (11) with the material dependent fractal dimension (12), whereas a set of plates of the same material with ratio $h/L = const$ is expected to obey the scaling relation (15) with the material dependent fractal dimension (18). Moreover, we expect that randomly folded plastic manifolds also obey the general scaling relation (21) with the material dependent scaling exponents defined by Eqs. (22) and (23).

## 3. Experiment details

Experiments with elastic sheets, such as latex rubber sheet (see [6]), requires measurements of the ball diameter under applied confinement force, because the sheet unfolds after withdrawing the confinement force. Unfortunately, in experiments with rubber strips under hydrostatic compression we were not able to prevent a deviation of ball form from a spherical shape. In experiments with (elasto-) plastic materials the deviations from spherical shape can be handled (see [100]). Early, it was noted that the geometrical and mechanical properties of randomly folded plastic and elasto-plastic materials belong to different universality classes [59]. Accordingly, in this work we performed the experiments with sets of randomly folded predominantly plastic aluminum foils of thickness $h = 0.02$ mm, earlier used in [59, 83], and elasto-plastic papers of thickness $h = 0.030 \pm 0.003$ and $0.068 \pm 0.005$ mm, earlier used in works [45, 46, 59].

Specifically, in this work we tested the sets of: 1) narrow strips ($L >> W = 2$ mm) of aluminum ($h = 0.02$ mm) and two papers of different thickness with the length varied according to the relation $L = \lambda L_0$ ($L_0 = 100$ mm) for scaling factor $\lambda = $ 1, 2, 3, 4, 5, 7.5, 9, 10, 12, 15, 24, 36, 48, and 50; 2) rectangle strips of length $L = 500$ mm with width varied from $W = W_0 = 1$ mm to $W = L$ with the relation $W = \lambda W_0$ for scaling factor $\lambda = $ 1, 2.5, 5, 25, 50, 100, 150, 200, 250, 500; 3) sets of square sheets of papers of thickness $h = 0.030 \pm 0.003$ and $0.068 \pm 0.005$ mm and size $2 \leq L \leq 100$ cm; and 4) paper strips of the same mass $M = \rho L W h$ ($h = const$, $\rho = 900$ kg/m$^3$) with the length to width ratio $X = L/W$ of 1, 4, 16, 64, 128, and 256 for $M/\rho h = 6400$ and $25600$. Furthermore, in this work we also used the data for: 5) sets of square sheets of aluminum foils of different thickness ($h = 0.02$, 0.06, 0.12, 0.24, and 0.32 mm) and size $60 \leq L \leq 600$ mm, early reported in [45, 46].

At least 20 strips of each geometry and size were folded in hands into approximately spherical balls. Further, to reduce the uncertainties caused by variations in the squeezing force the balls folded from aluminum foil strips were confined by applying the same force F=30 N along 15 directions taken at random (see [45]). In the case of randomly folded papers, to reduce uncertainties caused by strain relaxation after the folding force is withdrawn all measurements were performed ten days after the balls were folded, when no changes in the ball dimensions were observed (see [45, 46]). The diameter of each ball was defined as the average of measurements along 15 directions taken at random. Further, the mean diameter $R = R(L,W,h)$ was defined as the ensemble average over 20 balls folded from strips of the same mass and geometry.

## 4. Forced folding of predominantly plastic materials

Earlier, we reported that the square aluminum foils of different thickness obey the scaling behaviors (11) and (13) with the thickness independent fractal dimension $D_2(3) = 2.3 \pm 0.01$ and the scaling exponent $\phi = 0.35 \pm 0.01$. One can easily verify that the data reported in [83] are consistent with the scaling behavior (24) with the scaling exponents satisfying relations (12) and (17). Furthermore, from Eq. (18) with the experimental $\alpha = 0.21 \pm 0.02$ follows that $D_3(3) = 2.48 \pm 0.04$. Accordingly, Fig. 1 shows the data collapse in coordinates $f = M / \rho R^{D_3(3)}$ versus $X = h/L$ for folded square aluminum sheets of different thickness and edge size. The slope of fitting line $\theta = 0.15$ in Fig. 1 is consistent with the exponent $\theta = 0.16 \pm 0.05$ defined by Eq. (19).

In Fig 2 the mass $M = \rho L W h$ of randomly folded narrow aluminum strips ($W = 2$ mm $<< L$) of thickness of 0.02 mm is plotted versus the ball diameter $R$. On can see that the balls folded from narrow strips of the same thickness the experimental data obey the scaling behavior (20) with the fractal dimension $D_1(3) = 3.0 \pm 0.1$, close to the universal value $D_1(3) = 3$ predicted by the equality (9) for the one-dimensional elastic wires folded in three-dimensional space. So, our data suggest that the scaling properties of randomly folded narrow ($h < W << L$) plastic strips with $h/W = const$ are determined by the topological restrictions, analogous to those for one-dimensional elastic manifolds with finite bending rigidity.

Thus, the data of this work, together with the data from [83], suggest that the folded aluminum foils are characterized by the fractal dimensions $D_d(3)$ which are independent on the strip geometry characterized by the geometric constraint (2). Moreover, the data collapse shown in Fig. 3 in the coordinates $f = M / \rho R^{D_3(3)}$ versus $X = L h^\eta / W^{1+\eta}$ (with $\eta = 0.9$ defined by (22)) suggests that folded aluminum strips of different length, width, and thickness obey the scaling relation (21) with the fractal dimension $D_3(3) = 2.48 \pm 0.05$ defined by Eq. (18) and the scaling exponent $\vartheta = 0.17$ defined by Eq. (23).

It is interesting to note that the fractal dimension of folded three dimensional plates $D_3(3) = 2.48 \pm 0.05$ is close to the universal fractal dimension of the Apollonian sphere packing, $D_{AS}(3) = 2.4739465$ [112-114], which also coincides with the fractal dimension obtained by the analysis of the molecular surface volume of folded proteins [115, 116]. The free-volume distributions of folded proteins are broad, and the scaling of volume-to-surface and numbers of void versus volume show that the interiors of proteins are more like randomly packed spheres near their percolation threshold than like jigsaw puzzles. More deep analogy between the folded configurations and the packing of granular materials was discussed in [6, 116, 117, 118]. It was noted that in both cases the inherent states (stable configurations in the potential energy landscape) are distributed according to the principle of maximum Edwards's entropy (see also [119-121] and references therein). Accordingly, our finding may indicate that the folding configurations of randomly crumpled three-dimensional plates are determined by the maximization of Edwards's entropy. If so, the fractal dimension of the set of randomly crumpled plates (18), and hence the folding force scaling exponent, $\alpha = 3/D_3(3) - 1 = 0.21$, are determined by the Apollonian geometry of folded configurations.

## 5. Forced folding of elasto-plastic materials

The size of randomly folded elasto-plastic strips changes due to the strain relaxation after the confinement force is withdrawn (see [45]). As a result, the fractal dimension $D_2(3)$ of randomly folded sheets of the same thickness depends on the mechanical properties of material (see [45]). Nevertheless, in the case of randomly folded narrow paper strips ($W = 2$ mm $\ll L$) we found that the ball mass scales with ball diameter as

(20) with the same fractal dimension $D_1(3) = 3 \pm 0.1$ for two sets of strips made from papers of different thickness (see Fig. 2).

As it was pointed out above, the universal value of $D_1(3) = 3$ for elastic and plastic one-dimensional manifolds with the finite bending rigidity is determined by the topological restrictions leading to the condition (9). So, our data suggest that the increase in the ball diameters due to strain relaxation after withdrawing the confinement force does not change the scaling behavior (20) of randomly folded narrow elasto-plastic strips.

Figure 4 shows the graphs of paper ball mass versus its diameter for sets of hand folder paper strips of different geometry: a) sets of square sheets of thickness 0.03 and 0.069 mm and b) set of strips of length $L = 500$ mm and width $1 \leq W \leq 500$ mm from papers of thickness 0.03 mm and 0.068 mm. From graphs in Fig. 4a follows that the sets of folded square sheets of different thickness are characterized by different fractal dimensions. Namely, we found that $D_2(3) = 2.25 \pm 0.05$ and $D_2(3) = 2.54 \pm 0.06$ for folded papers of thickness 0.03 mm and 0.068 mm, respectively; in agreement with the corresponding values reported in [83]. The slopes of graphs in Fig. 4b are consistent with the material (thickness) dependent exponent

$$\mu = \frac{3D_2(3)}{6 - D_2(3)} \qquad (25)$$

expected from the scaling behavior (1) with the scaling function (7) of the geometric constraint $X = W/L$ and the scaling exponent (8). In Fig. 5 of diameter of balls folded from the strips of the fixed mass ($M/\rho = 6400$ mm$^2$) versus the geometric constraint $X = W/L$ for paper two papers of different thickness. The slopes of straight lines in Fig. 5 are consistent with the material (thickness) dependent exponent

$$\upsilon = \frac{1}{D_2(3)} - \frac{1}{3}, \qquad (26)$$

obtained from the scaling behavior (1) with the scaling function (7) of the geometric constraint $X = W/L$ and the scaling exponent (8).

Furthermore, from graphs in Fig. 6 it follows that sets of randomly folded paper strips obey the scaling relation (1) with the scaling function (7) of the geometric constraint $X = W/L$ and the thickness (and material) dependent scaling exponent $\theta$ defined by Eq. (8). Thus, for balls folded from strips of the same paper ($h = const$) there is a topological crossover between the set of randomly folded two-dimensional sheets obeying the fractal behavior (11) with the material dependent fractal dimension $D_2(3)$ to a set of randomly packed strings obeying the fractal behavior (20) with the universal fractal dimension $D_1(3) = 3$. However, folded paper strips do not obey the scaling relation (15) because of the strain relaxation rate depends on the paper thickness. So, for sets of randomly folded elasto-plastic manifolds after relaxation the scaling relation (21) is not valid.

## 6. Conclusions

Experimental data suggest that randomly folded plastic manifolds obey the scaling behavior (21)-(23) with the geometry independent fractal dimensions $D_d(3)$ defined by Eqs. (9), (12), and (18). Accordingly, the force-diameter relation for square plastic sheets (24) can be generalized as

$$\frac{R}{L} \propto \left(\frac{W}{h}\right)^{2\beta} \left(\frac{W}{L}\right)^{2/3} \left(\frac{Kh^2}{FW}\right)^{\alpha}, \qquad (27)$$

such that folded manifolds obey the fractal scaling behavior (21)-(23). Hence, the force-diameter relation (27) implies the topological crossovers from the folding of three-dimensional plates ($h/L = const << W/L = const \leq 1$) obeying the fractal law (15) to the folding of two-dimensional sheets ($h = const << W$, $W/L = const \leq 1$) obeying the fractal law (11) and further to the packing of one-dimensional strings ($h = const \leq W = const << L$) obeying the fractal law (20).

In the case of ideally elastic manifolds with the bending rigidity $\kappa \propto Eh^3$, analogous considerations lead to the following force-diameter relation

$$\frac{R}{L} \propto \left(\frac{W}{h}\right)^{2\beta} \left(\frac{W}{L}\right)^{2/3} \left(\frac{Eh^3}{FW}\right)^{\alpha}, \qquad (28)$$

which generalizes the force-diameter relation (10) to the case of elastic plates with $h << W \leq L$. So, the scaling relation (28) can be easily verified by numerical simulations with a coarse-grained model of triangulated surfaces with bending and stretching elasticity (see Ref. [5]).

We pointed out that in the both cases, the force scaling exponent $\alpha$ is independent of the topological dimension $d$ and the strip geometry. Furthermore, we noted that the density of folding energy ($U = PV \propto (F/R^2)R^3$) per unit volume of square sheet ($hL^2$) decreases as $U/hL^2 \propto L^{(2/D_2(3)-2)}$ when the sheet size increases.

In the case of elasto-plastic materials, such as paper sheets, the scaling force-diameter relation (28) is expected to be also valid during the (fast) folding under increasing confinement force $F$. However, after the confinement force is withdrawn, a slow strain relaxation leads to the increase of $R$, such that a set of randomly folded sheets of the same thickness is characterized by the material and thickness dependent fractal dimension $D_2(3)$, whereas sets of randomly folded strings ($h = const \leq W = const << L$) are characterized by the universal fractal dimension $D_1(3) = 3$ before and after stress relaxation. Nevertheless, we found that folded elasto-plastic strips of the same thickness obey the scaling relation (1) with the scaling function (7) of the geometric constraint $X = W/L$ with the material dependent scaling exponent (8), implying a continuous topological crossover from folding of two-dimensional sheets to packing of one-dimensional strings. This finding suggests that the rate of strain relaxation after withdrawing the confinement force depends on the $X = W/L$.

It is interesting to note that the diameter of sheets folded under the same pressure ($P \propto F/R^2 = const$) scales with the sheet size as $R \propto L^\nu$, where $\nu = (1+2\beta-\alpha)/(1-2\alpha)$. Hence a set of sheets folded under fixed pressure obeys a fractal law $M \propto R^D$, where $D = 2(1-2\alpha)/(1+2\beta-\alpha) < 2$. It is obvious that the internal configuration of a folded sheet does not know the way of the set forming and is characterized by another fractal dimension $D_l \geq D_2(3) > D$. This leads to the intrinsically anomalous self-similarity of the set of folded sheets, as studied in [46]. Moreover, it was found that the internal configurations of folded sheets are characterized by the universal local fractal dimension $D_l = 2.64 \pm 0.05$ [46].

## Acknowledgements

We wish to thank Professor T. A. Witten for the comments concerning the density of folding energy stored in the fractal balls. This work was supported by the Government of Mexico City under Project PICCT08-38.

**Figure captions:**

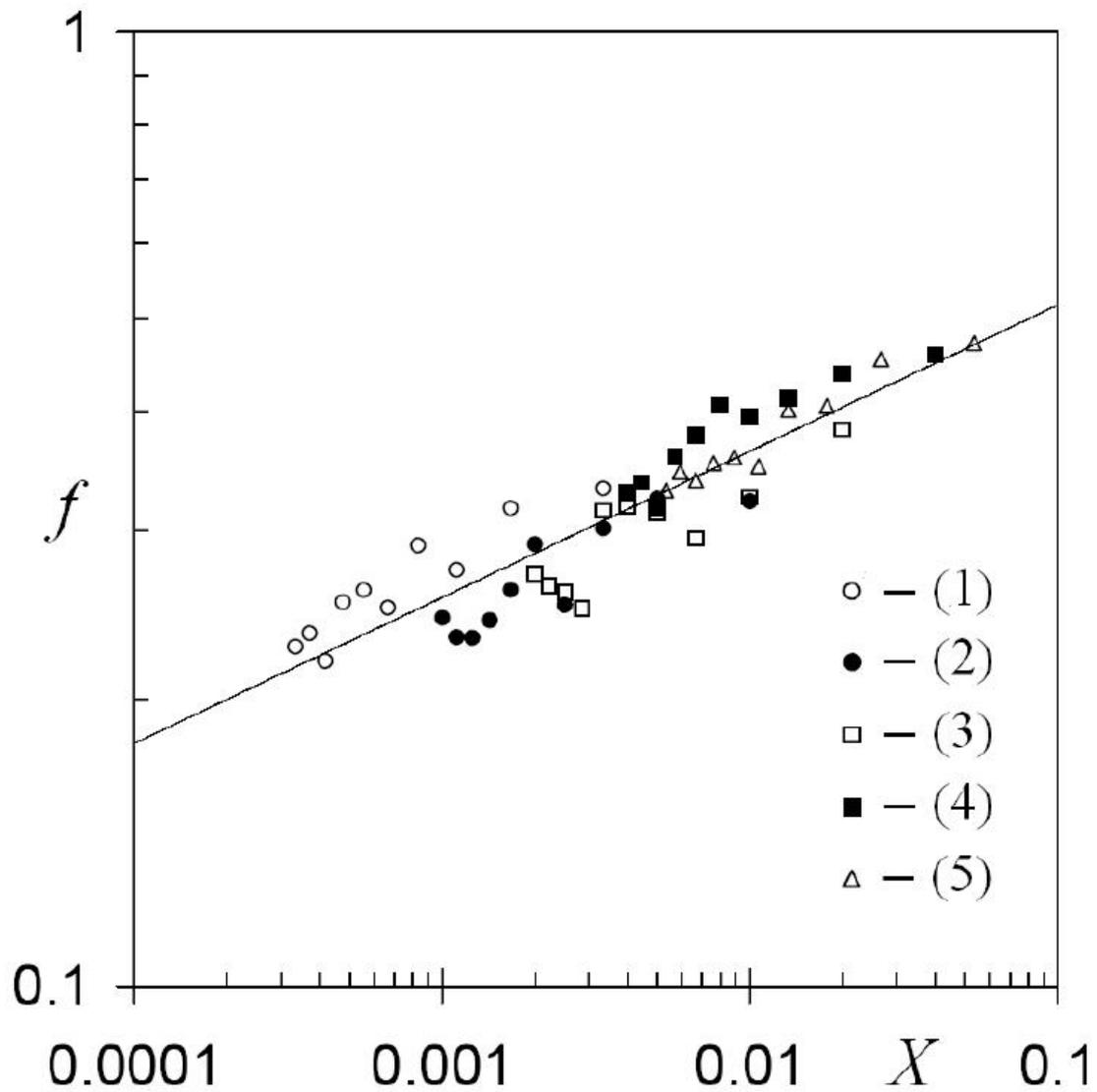

**Figure 1.** Data collapse in coordinates $f = M/\rho R^{2.48}$ (mm$^{0.52}$) versus $X = h/L$ (arbitrary units) for randomly folded foils ($W = L$) of thickness $h = 0.02$ (1), 0.06 (2), 0.12 (3), 0.24 (4), and 0.32 mm (5) studied in the work [83]. The slope of straight line is 0.15.

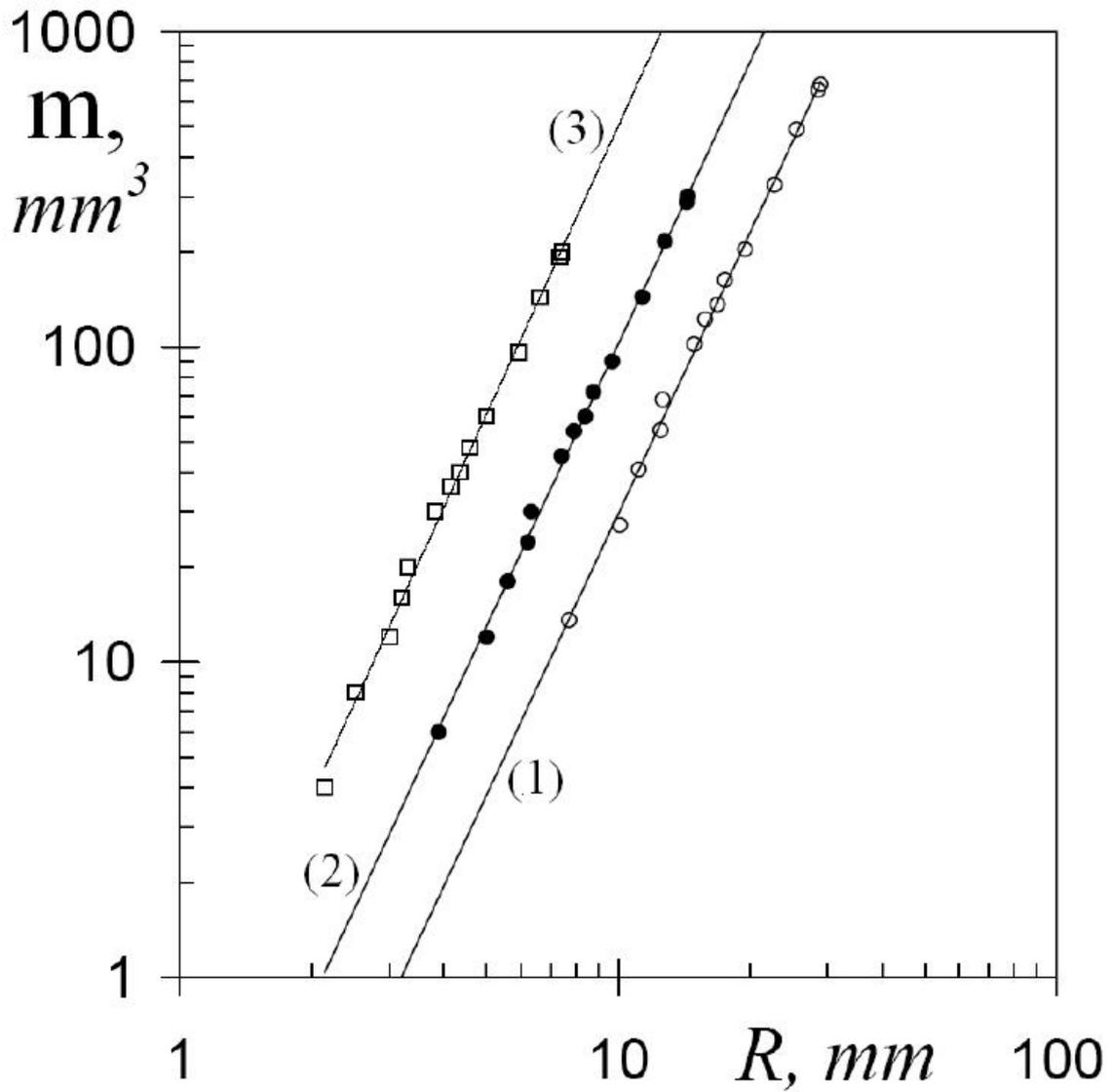

**Figure 2**. Log-log plots of $m = M/\rho = LWh$ (mm$^3$) versus $R$ (mm) for sets of randomly folded strips of width $W = 2$ mm from papers of thickness 0.068 (1) and 0.03 mm (2) and aluminum of thickness 0.02 mm (3). The slopes of all fitting lines are equal to $D_1(3) = 3 \pm 0.1$.

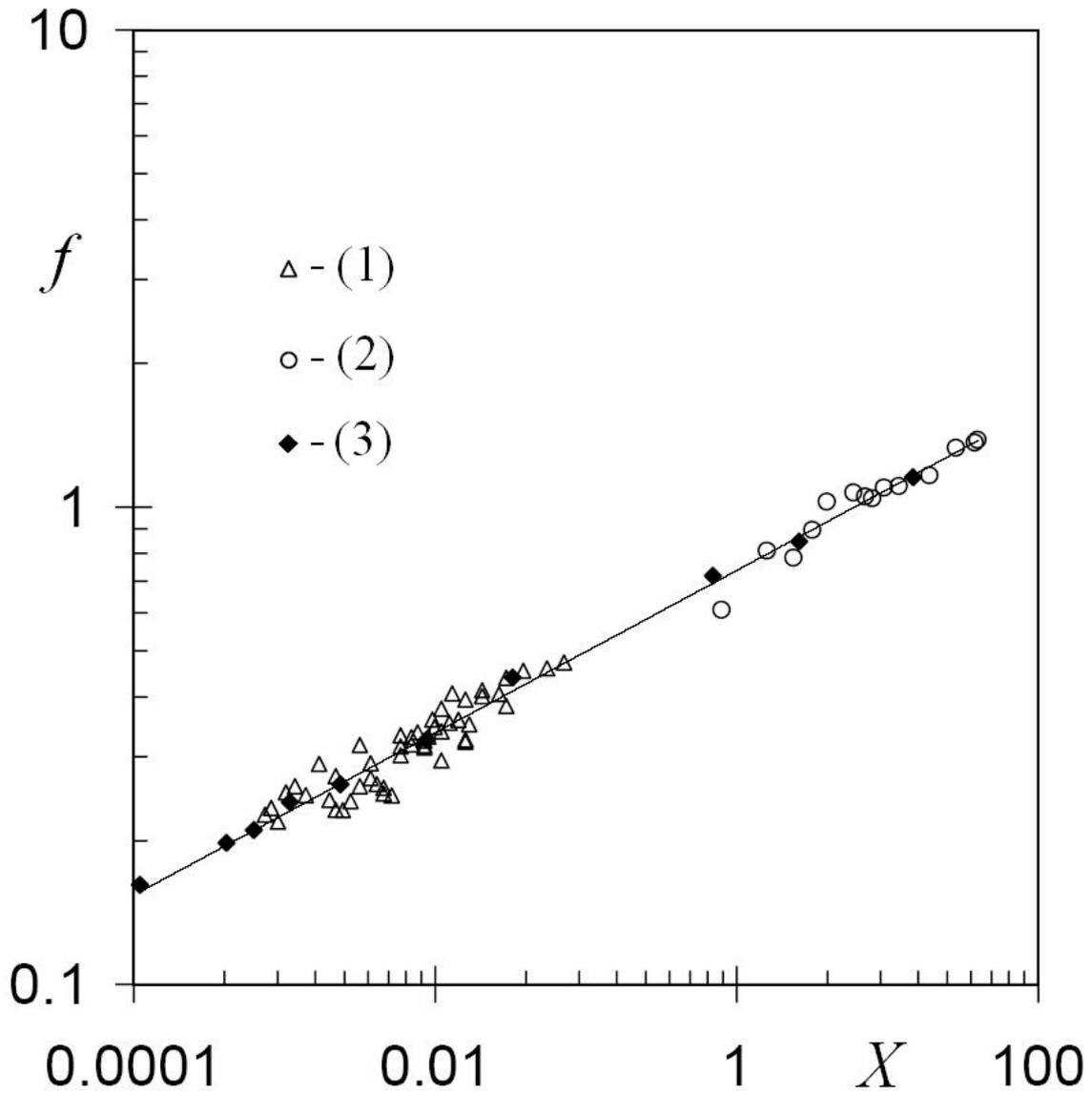

**Figure 3**. Data collapse in coordinates $f = M/\rho R^{2.48}$ (in arbitrary units) versus the geometric constrain $X = Lh^{0.9}/W^{1.9}$ for randomly folded aluminum foils of different geometry: triangles – the square sheets (the data from [83], the same that used in Fig. 1), circles – the strips of width $W = 2$ mm from aluminum of thickness 0.02 mm (see Fig. 2), and rhombs – the strips of length $L = 500$ mm. The slope of fitting line is 0.17.

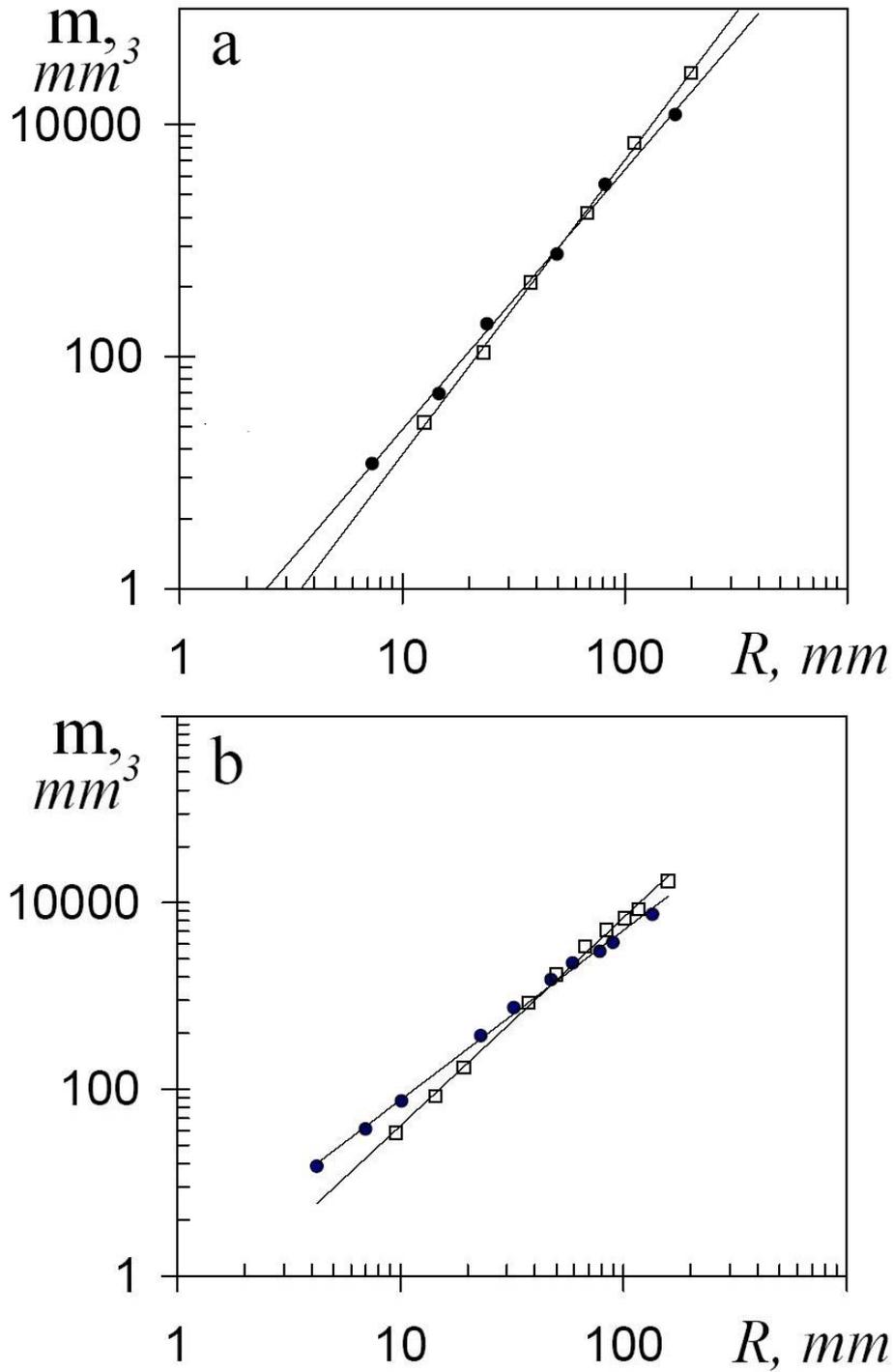

**Figure 4**. Log-log plots of $m = M/\rho = LWh$ (mm$^3$) versus $R$ (mm) for sets of randomly folded papers: a) sets of square sheets of thickness 0.03 (full circles) and 0.069 mm (squares); the slopes of fitting lines are $D_2(3) = 2.25$ and $D_2(3) = 2.54$, respectively; b) set of strips of length $L = 500$ mm from papers of thickness 0.03 mm (full circles) and 0.068 mm (squares); the slopes of fitting lines are 1.8 and 2.2, respectively.

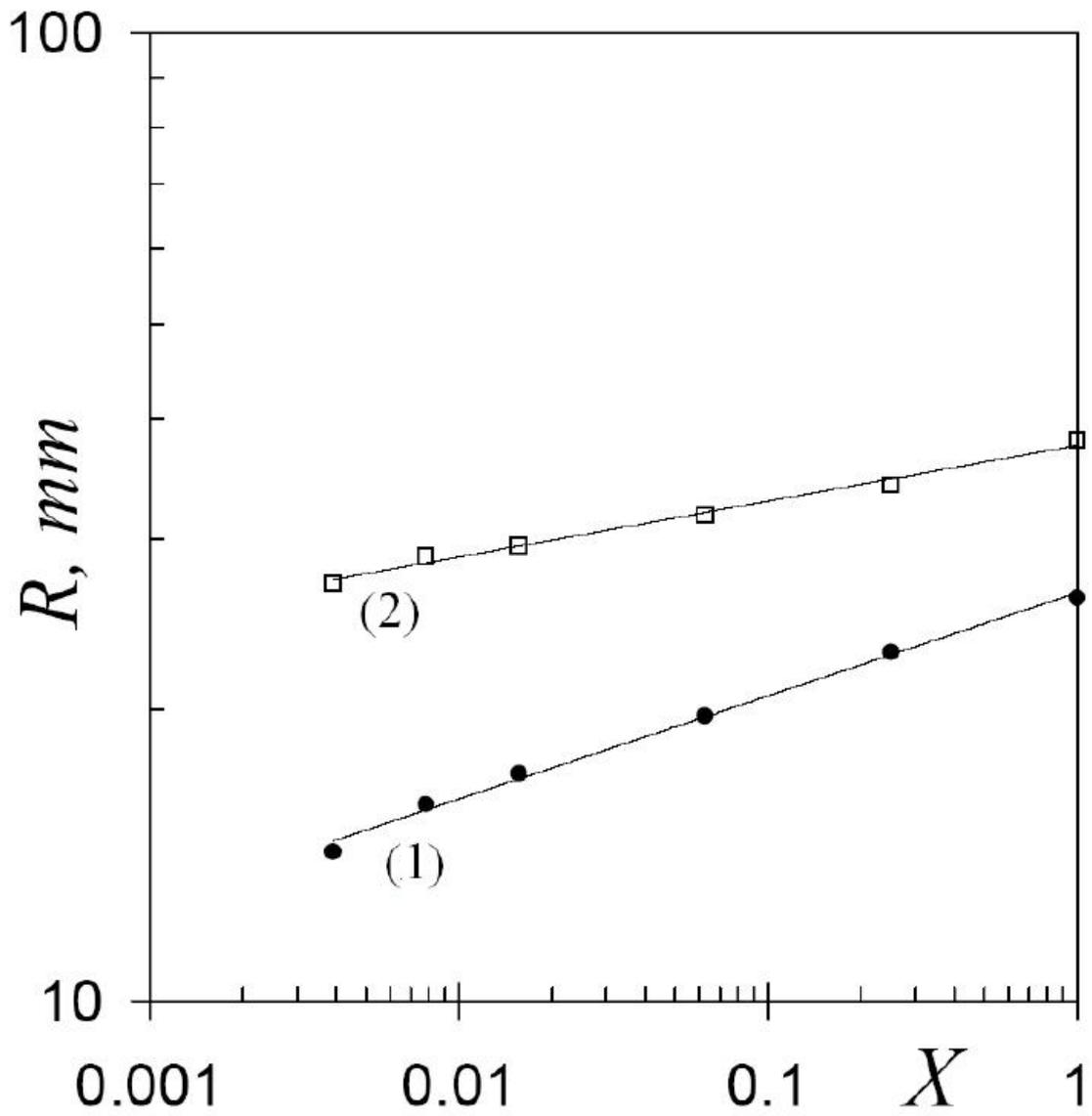

**Figure 5**. The ball diameter $R$ (mm) versus the geometric constrain $X = W/L$ for the sets of randomly folded paper sheets of the mass $M/\rho = 6400$ mm$^2$ with the thickness $h = 0.03$ mm (1) and 0.068 mm (2); the slopes of straight lines are 0.12 (1) and 0.06 (2), respectively.

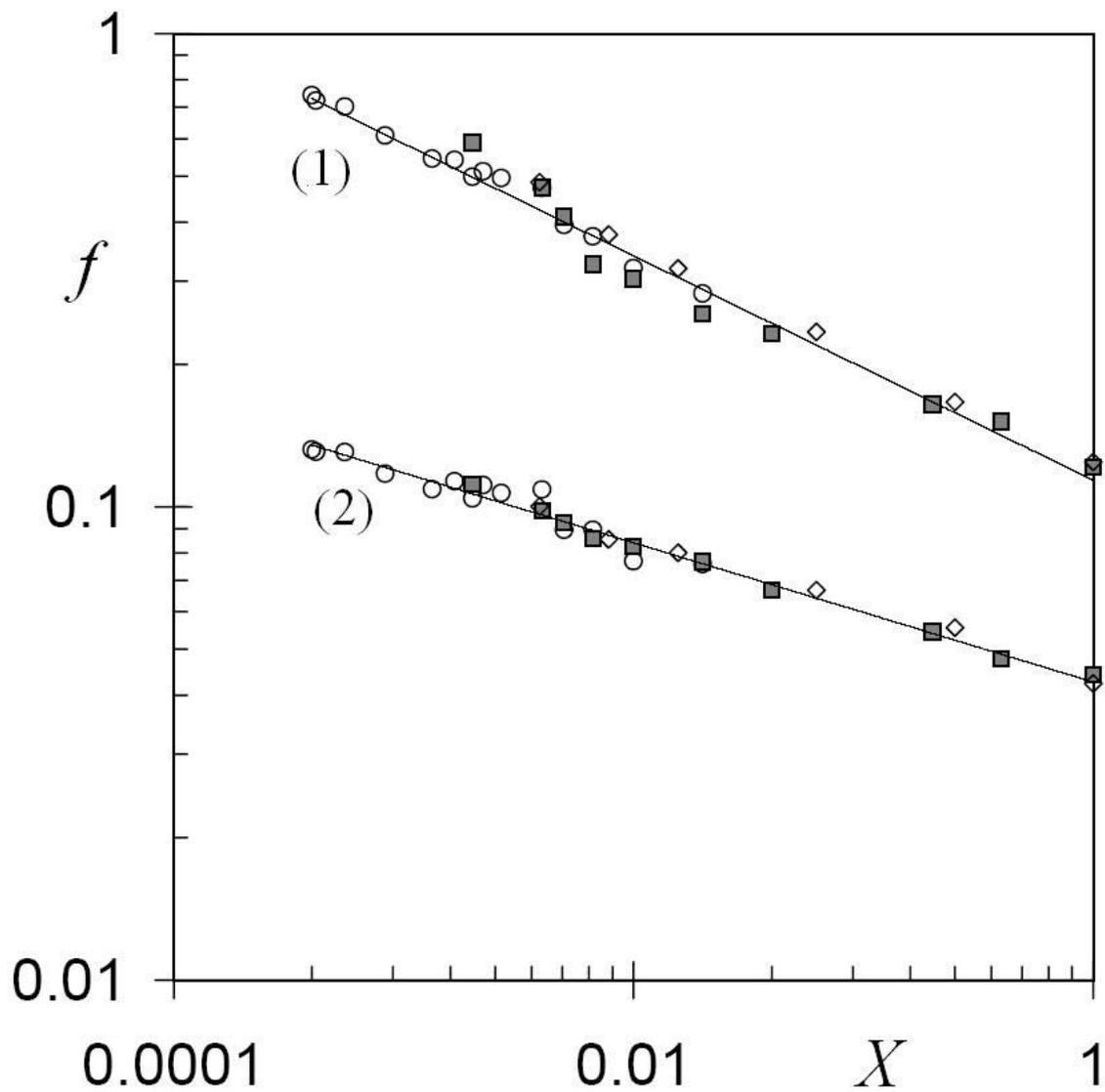

**Figure 6.** Data collapse in coordinates $f = M/\rho R^{D_2}$ (in arbitrary units) versus $X = W/L$ for randomly folded sheets of paper of different geometries with the thickness $h = 0.03$ mm (1) and 0.069 mm (2): circles – the strips of width 2 mm (see Fig. 2), full squares – the strips of length 500 mm (see Fig. 4 b), and rhombs – the strips with $L \times W = const$ (see Fig. 5). The slopes of straight lines are $D_2/D_1 - 1 = $ -0.24 (1) and -0.15 (2).